\documentclass[lettersize,journal]{IEEEtran}
\usepackage{amsmath,amsfonts}
\usepackage{algorithmic}
\usepackage{array}
\usepackage[caption=false,font=normalsize,labelfont=sf,textfont=sf]{subfig}
\usepackage{textcomp}
\usepackage{stfloats}
\usepackage{url}
\usepackage{verbatim}
\usepackage{graphicx}
\def\BibTeX{{\rm B\kern-.05em{\sc i\kern-.025em b}\kern-.08em
    T\kern-.1667em\lower.7ex\hbox{E}\kern-.125emX}}
\usepackage{balance}
\ifCLASSINFOpdf
\else
   \usepackage[dvips]{graphicx}
\fi
\usepackage{tabularray, amsmath, bm, color, multirow}
\usepackage{booktabs}

\usepackage[table]{xcolor}
\usepackage{caption}
\usepackage{adjustbox}
\usepackage{graphicx, hyperref}

\begin{document}

\title{Melody-Lyrics Matching with Contrastive Alignment Loss}

\author{Changhong Wang, Michel Olvera, Gaël Richard, \IEEEmembership{Fellow, IEEE}
\thanks{This work has been submitted to the IEEE for possible publication. Copyright may be transferred without notice, after which this version may no longer be accessible.}
\thanks{This work was funded by the European Union (ERC, HI-Audio, 101052978). Views and opinions expressed are however those of the author(s) only and do not necessarily reflect those of the European Union or the European Research Council. Neither the European Union nor the granting authority can be held responsible for them. }
\thanks{The authors are with the Laboratoire de Traitement et Communication de l’Information (LTCI), Télécom Paris, Institut Polytechnique de Paris, 91120
Palaiseau, France (e-mail: \{changhong.wang; olvera; gael.richard\}@telecom-paris.fr).}}


\maketitle

\begin{abstract}
The connection between music and lyrics is far beyond semantic bonds. Conceptual pairs in the two modalities such as rhythm and rhyme, note duration and syllabic stress, and structure correspondence, raise a compelling yet seldom-explored direction in the field of music information retrieval.
In this paper, we present \emph{melody-lyrics matching} (MLM), a new task which retrieves potential lyrics for a given symbolic melody from text sources.  
Rather than generating lyrics from scratch, MLM essentially exploits the relationships between melody and lyrics.
We propose a self-supervised representation learning framework with contrastive alignment loss for melody and lyrics.
This has the potential to leverage the abundance of existing songs with paired melody and lyrics. 
No alignment annotations are required.
Additionally, we introduce \emph{sylphone}, a novel representation for lyrics at syllable-level activated by phoneme identity and vowel stress. 
We demonstrate that our method can match melody with coherent and singable lyrics with empirical results and intuitive examples. 
We open source code and provide matching examples on the companion webpage:  \texttt{\url{https://github.com/changhongw/mlm}}.
\end{abstract}


\IEEEpeerreviewmaketitle

\section{Introduction}

\IEEEPARstart{A}{longside} advancements in music generation, there has been a growing interest in lyrics or song generation in the music information retrieval community. Generating lyrics from scratch is non-trivial, as it requires considerations of not only text quality, but also the relationships between music and lyrics. 
Prior efforts have sought to address both aspects simultaneously~\cite{watanabe2018melody, ma2021ai}, but have encountered difficulties, primarily from a lexical standpoint, including grammatical correctness and semantic consistency. 
These works also relied on a strong assumption of syllabic singing, meaning each melody note corresponds to a single syllable~\cite{pan2022vocal}.
However, this assumption does not always hold true, for example, in cases of melisma, where multiple notes are sung to a single syllable~\cite{panteli2017towards}.
To date, no research work has created lyrics solely from the second perspective, i.e., exploring and leveraging the relationships between music and lyrics to match music with existing texts.

\begin{figure}[t]
    \centering   
    \includegraphics[width=\linewidth]{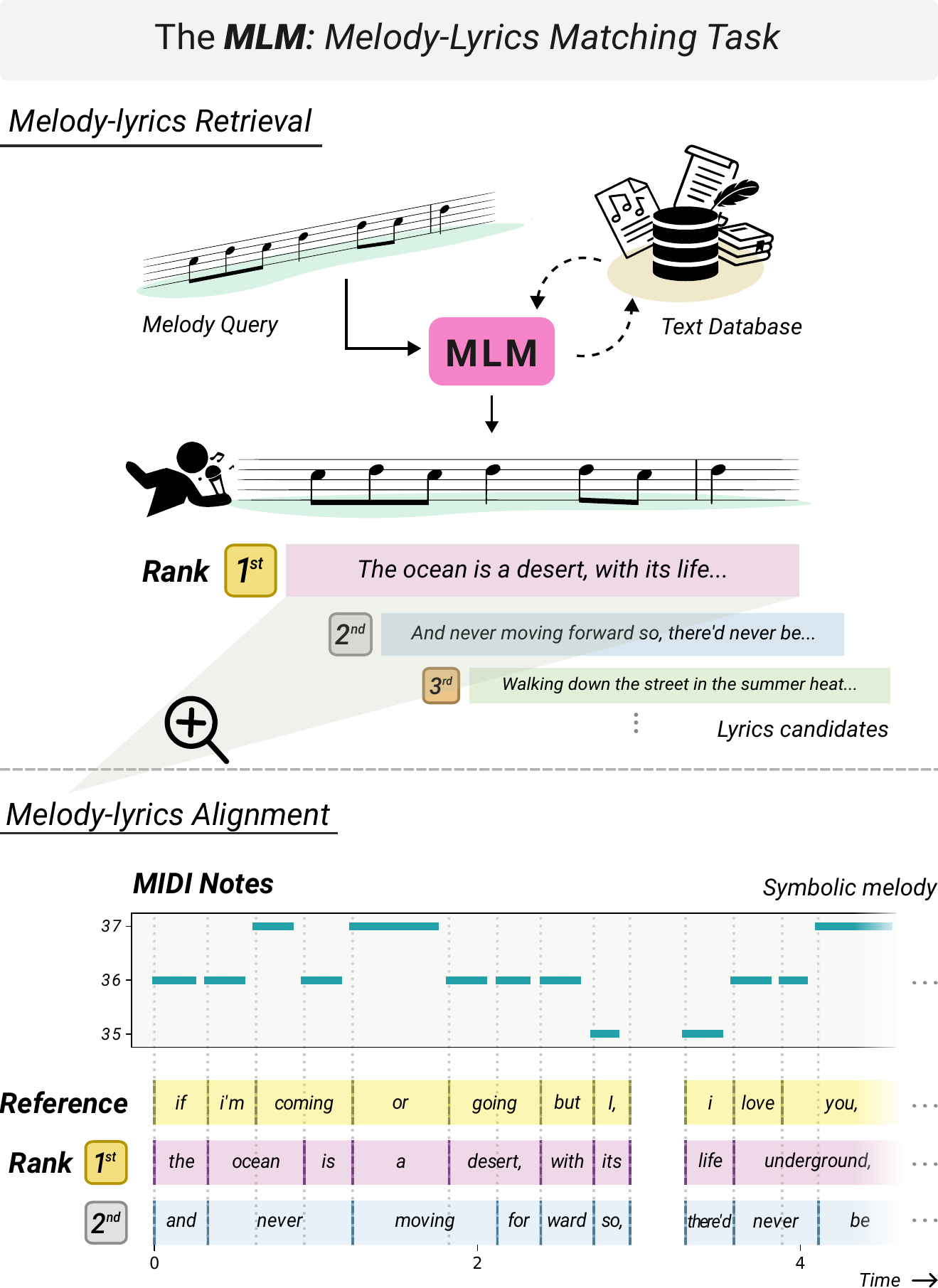}
    \caption{Given a melody query, melody-lyrics matching (MLM) retrieves lyrics from a text database and returns a ranked list of singable candidates whose words and syllables (see Fig.~\ref{fig:example_match}) are aligned with the melody notes. MLM performs retrieval and alignment jointly.}
    \label{fig:MLM_task}
\end{figure}

From this fresh angle, we propose a new task, called \emph{melody-lyrics matching} (MLM), which retrieves potential lyrics for a symbolic melody query from existing text sources, as shown in Fig.~\ref{fig:MLM_task}.
Given a melody query, either existing or newly-composed, MLM searches a text database for lyrics and returns a ranked list of lyrics candidates.
It also provides note-syllable alignment between the melody and lyrics so that users can sing along directly with the melody.
The motivation for MLM stems from the abundance of text resources that could serve as lyrics candidates such as poems, verses, lyrics from other songs, and general text in a broader sense.
These texts are typically grammatically correct, coherent, structured, and possibly rhymed.
Matching melody with such texts allows us to focus primarily on the relationships between music and lyrics, rather than on the challenges associated with natural language processing (NLP). 
MLM can enrich our singing experience as one can match a single melody to various lyrics or align the same lyrics to different melodies.

Despite the lack of prior work, MLM closely resembles the lyrics retrieval task~\cite{kruspe2016retrieval}, which aims to find the original lyrics of a song from a lyrics database.
The key difference is that MLM can fetch suitable lyrics not only for existing songs but also for newly composed melodies.
Moreover, it provides alignment between notes and syllables, rather than plain-text lyrics alone.
Another related task is audio-to-lyrics alignment~\cite{durand2023contrastive, schulze2021phoneme, vaglio2020multilingual}, which synchronizes the audio of singing with the corresponding texts.
Again, this alignment is only feasible between the singing voice and its corresponding lyrics, which are connected by an acoustic model. 
In contrast, for MLM, the input is only a symbolic melody, which does not contain explicit information about the lyrics.
This weak connection makes it possible to match different lyrics to the same melody, forming the key idea behind MLM.

In a broader context, MLM  can be viewed as a form of \emph{alignable sequence retrieval}, which aims to retrieve all potential sequences that can be aligned with a given sequence query, whether in the same or different modalities. 
Similar tasks have been explored in the literature, including 
music recommendation for movie videos~\cite{pretet2022video}, text-music retrieval~\cite{komatsu2025aligned}, identification of alignable videos from large video databases~\cite{dave2025sync}, sequential contrastive audio-visual learning~\cite{tsiamas2025sequential} and
video-text representation learning~\cite{ko2022video}.
Compared to existing frame-wise contrastive learning approaches~\cite{wu2025flam}, MLM is more challenging due to the loose correspondence between melody and lyrics, as well as the requirement to provide explicit alignment.
Additionally, there is not yet any work on lyrics retrieval from this perspective.

We approach MLM as a general self-supervised representation learning problem.
Our method learns a dual encoder for melody and lyrics using a contrastive alignment loss, as illustrated in Fig.~\ref{fig:MLM}.
The framework leverages the abundance of existing songs with paired music and lyrics for training. 
No alignment annotations are required.
It enables the retrieval of potential lyrics for a given melody query, without relying on any language model. 

We employ the soft dynamic time warping (SDTW) loss~\cite{cuturi2017soft} for alignment, as it is differentiable and takes into account all possible alignment paths, rather than only the optimal one returned by classical dynamic time warping (DTW)~\cite{muller2015fundamentals}.
This characteristic is desirable for MLM, as it aims to explore lyrics that can be plausibly paired with the same melody. 
In summary, we present the following contributions:
\begin{itemize}
    \item \textit{Melody-lyrics matching (MLM)}: a new task that retrieves potential lyrics for a given melody query from a text database. 
    \item \textit{Sylphone representation}: a novel representation for lyrics that captures the relationships between melody and lyrics at the syllable level.
    \item Self-supervised representation learning for MLM with contrastive alignment loss, requiring no alignment annotations. 
    \item A benchmark dataset as well as a set of objective metrics for evaluating MLM.
\end{itemize} 

\section{MLM with Contrastive Alignment loss}

\subsection{Melody-Lyrics Matching Task}\label{sec:task}
\begin{figure}[t]
    \centering   
    \includegraphics[width=0.9\linewidth]{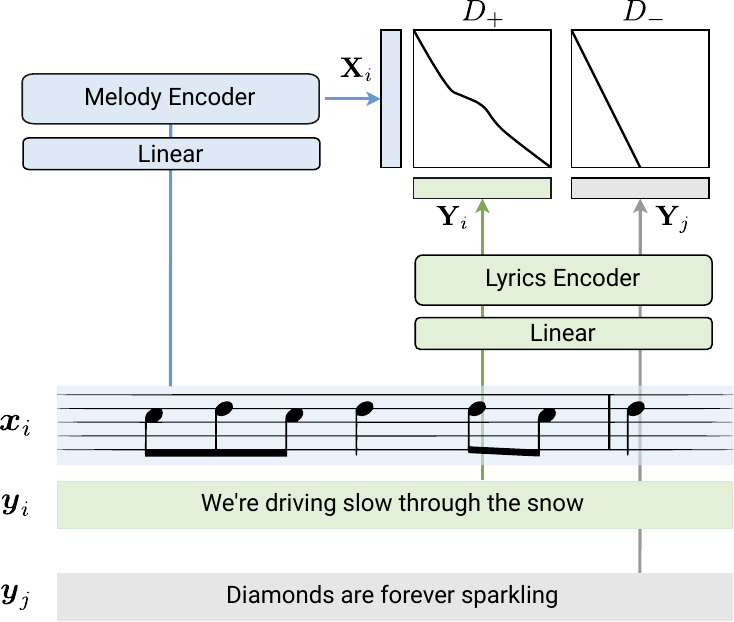}
    \caption{Melody-lyrics matching (MLM) with contrastive alignment loss. Given a melody query ${\bm x}_i$, the system retrieves potential lyrics ${\bm y}$ from a text database. It learns a dual encoder for melody and lyrics by minimizing the alignment cost $D_+$ between positive embedding pairs $(\mathbf{X}_i, \mathbf{Y}_i)$ from the same song,  while maximizing $D_-$ for negative pairs $(\mathbf{X}_i, \mathbf{Y}_j)$ from different songs. No additional annotation is required beyond lyrics from existing songs.}
    \label{fig:MLM}
\end{figure}

Given a melody query $\bm{x}$, the melody-lyrics matching (MLM) task aims to find all lyrics $\bm{y}$ from a text database that can be sung along with $\bm{x}$. 
Fig.~\ref{fig:MLM} illustrates our proposed method, a dual-encoder framework with contrastive alignment loss (MLM-CAL).
The encoders map the melody and lyrics candidates into embeddings $\mathbf{X}$ and $\mathbf{Y}$, respectively. 
A pair $(\mathbf{X}_i, \mathbf{Y}_i)$ is considered positive when the melody and lyrics come from the same song, whereas $(\mathbf{X}_i, \mathbf{Y}_j)$ forms a negative pair when this is not the case.
Note that both $\mathbf{X}$ and $\mathbf{Y}$ are sequences of embedding vectors for each example, instead of single embedding vectors as in conventional contrastive learning~\cite{elizalde2023clap}. 

Using the SDTW loss $D_+$ of positive melody-lyrics pairs alone is one possible approach to training the model in Fig.~\ref{fig:MLM}.
However, this learning scheme may collapse to trivial solutions~\cite{ko2022video}. 
To address this, we incorporate information from unmatched pairs (i.e. negative pairs) through contrastive learning.
Specifically, we minimize the alignment cost $D_+$ for positive pairs while maximizing $D_-$ for negative pairs.

For melody and lyrics input, we use note- and syllable-level features, respectively. 
In this work, we focus solely on melodic information for melody sequences, including pitch, note duration, and note onset shift.
Through the following quantization scheme, we construct a 177-dimensional feature vector for each note. 

\begin{itemize}
    \item \textbf{Pitch change} (129-D): Normalized to the first note, this feature includes a 128-dimensional one-hot vector to encode MIDI pitch change, along with a binary value indicating the sign of the pitch change. 
    \item \textbf{Duration} (24-D): Represented by $\log (t)$, where $t$ is the note duration in seconds. This value is normalized to the range $[0,1]$, and then quantized into a 24-dimensional one-hot vector.
    \item \textbf{Onset shift} (24-D): Processed similarly to duration, onset shifts are defined as inter-onset intervals in seconds.
\end{itemize}
Pitch change normalization is motivated by the important role that pitch contour plays in the relationship between melody and lyrics~\cite{nichols2009relationships}.
Quantizing note duration and onset shift helps smooth out noise~\cite{liu2022performance}.

\subsection{Sylphone Representation}\label{sec:sylphone}

Existing literature on lyrics representations primarily emphasizes semantic content through token-level text embeddings \cite{ma2021ai, durand2023contrastive, liangxai}, which overlook the acoustic relationships between melody and lyrics.
In contrast, we propose a new representation for lyrics that incorporates phonetic features.
Phonetic connections play a crucial role in the relationships between melody and lyrics~\cite{park2024real}, as evidenced by the phenomenon of \emph{mondegreen} in lyrics perception.
This occurs when a listener misinterprets lyrics, often perceiving a word or phrase that differs from the original but shares a similar pronunciation.
For example, \textit{``kiss the sky"} is commonly misheard as \textit{``kiss this guy"} in \emph{Purple Haze} by Jimi Hendrix.  
The concept of mondegreen supports the possibility of fetching different lyrics for the same melody, which are acoustically connected to the melody but may vary in semantic meaning or lexical format.

We propose \emph{sylphone}, a novel representation for lyrics that encodes each syllable as a multi-hot vector activated by phoneme identity and vowel stress. 
This design is motivated by the observation that lyrics typically align with melody notes at the syllable level~\cite{low2003singable, antonisen2024polysinger}.
To process a lyrics sequence, we first obtain the corresponding phonemes for each word using the CMU Pronouncing Dictionary\footnote{http://www.speech.cs.cmu.edu/cgi-bin/cmudict}, which employs the ARPABET notation for representing English phonemes.
Besides phoneme transcriptions, the dictionary optionally provides lexical stress annotations for vowels, with three values: 0, 1, 2, indicating increasing degrees of stress.
We keep the stress levels for vowels, group phonemes by syllable, and derive syllable-level phonemes, referred to as \emph{sylphones}.
Each sylphone consists of one vowel, one stress level, and one or more 
consonants.
Table~\ref{tab:sylphone} presents the sylphone sequence for a lyrics example taken from the song \emph{I Hate This Part} by The Pussycat Dolls in the DALI dataset (see Section~\ref{sec:dataset}). In this example, the sylphone representation of \textit{``We're''} is [W, \textbf{IY1}, R], where W and R are consonants, \textbf{IY} is a vowel and \textbf{1} indicates its associated stress level.



Another important design of the sylphone representation is its distinctions between consonants occurring before and after the vowel, categorizing them as \emph{front constants} and \emph{end consonants}, respectively.
This distinction helps capture rhyme information.
Rhyme refers to syllables sharing the same vowel~\cite{pattison1991songwriting}.
There are different rhyme categories depending on the similarity between the end consonants.
Specifically, with a shared rhyming vowel, greater phonological similarity among end consonants leads to a stronger rhyme.
For instance, a perfect rhyme occurs when syllables share the exact same vowel and end consonants, if any.
Subsequently, we refer to the combination of the vowel and end consonants of a sylphone as its \emph{rhyming elements}.

\begin{figure}[t]
    \centering
    \includegraphics[width=\linewidth]{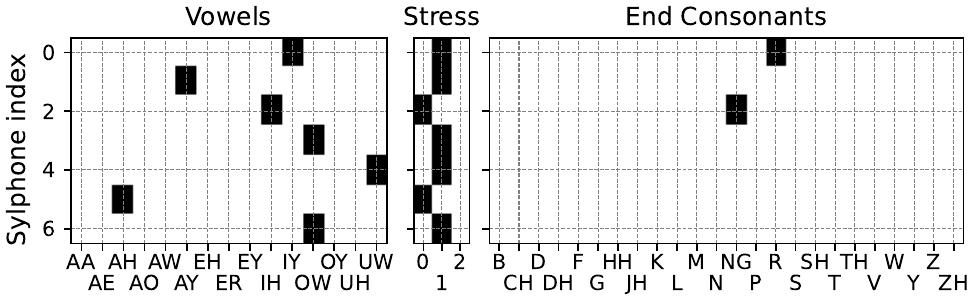}
    \caption{Visualization of sylphone representation of lyrics in Table~\ref{tab:sylphone}. X-axis labels are vertically staggered to enhance readability.}
    \label{fig:sylphone_visual}
\end{figure}

To focus on the core connection between melody and lyrics, we consider only the rhyming elements and stress of sylphones in this work. We expand each component—vowel, stress, and end consonant(s)—into a one- or multi-hot vector, followed by concatenation. 
Following the International Phonetic Alphabet (IPA) system, which has 15 vowels, 3 stress levels, and 24 consonants, our sylphone representation results in a 42-dimensional vector (15 + 3 + 24).
Fig.~\ref{fig:sylphone_visual} displays the sylphone representation for the lyrics example presented in Table~\ref{tab:sylphone}.
According to~\cite{nichols2009relationships}, stopword (non-functional words) rarely corresponds to long notes in melodies.
Motivated by this fact, we also include a binary value to indicate whether the sylphone originates from a stopword\footnote{We use the English stopword list from the Natural Language Toolkit: https://www.nltk.org.} \cite{bird2009natural}.  
This additional feature results in an input representation of 43 dimensions for the lyrics encoder in Fig.~\ref{fig:MLM}.

\begin{table}[t]
\centering
\caption{Sylphone sequence for a lyrics example}
\begin{adjustbox}{center}
\begin{tabular}{@{}lccccccc@{}}
\toprule
\rowcolor{gray!10}
\textbf{Syllable} & We're & dri & ving & slow & through & the & snow \\
\midrule
 & W     & D, R & V    & S, L & TH, R   & DH  & S, N \\
\textbf{Sylphone} & \textbf{IY1} & \textbf{AY1} & \textbf{IH0} & \textbf{OW1} & \textbf{UW1} & \textbf{AH0} & \textbf{OW1} \\
 & R     & /     & NG   & /     & /       & /   & / \\
\bottomrule
\end{tabular}
\end{adjustbox}
\label{tab:sylphone}
\end{table}

\subsection{Soft Dynamic Time Warping Loss}
As MLM requires alignment between melody and lyrics, we employ a differentiable alignment loss. 
Soft Dynamic Time Warping (SDTW)~\cite{cuturi2017soft} serves this purpose by replacing the $\min$ operation of classical Dynamic Time Warping (DTW) with a soft version, thereby achieving differentiability. 
Let $A$ denote the alignment path matrix for a melody-lyrics pair $(\mathbf{X},\mathbf{Y})$, where $\mathbf{X}$ has $n$ notes and $\mathbf{Y}$ contains $m$ sylphones.
$\mathcal{A}_{n,m} \subset \{0,1\}^{n \times m}$ denotes the set of all valid path matrices. 
With the pairwise distance matrix $C(\mathbf{X},\mathbf{Y})$, the goal of the classical DTW is to find the path with the minimum accumulative cost:
\begin{equation}
    D(\mathbf{X},\mathbf{Y}) = \min_{A \in \mathcal{A}(n,m)} <A, C(\mathbf{X},\mathbf{Y})>.
\end{equation}
This corresponds to the sum of the pairwise cost along the optimal alignment path.
Due to the non-differentiability of the $\min$ operation, an approximation replacement was considered in~\cite{cuturi2017soft}:
\begin{equation}
    {\min}^{\gamma}\{a_1, ..., a_n\} := 
    \begin{cases}
      \min_{i\leq n} a_i & \gamma = 0\\
      -\gamma \log \sum_{}^{} \exp(-a_i/\gamma) & \gamma > 0,
    \end{cases}   
\end{equation}
which leads to the soft-DTW (SDTW):
\begin{equation}
    D^{\gamma}(\mathbf{X},\mathbf{Y}) := {\min}^{\gamma} \{<A, C(\mathbf{X},\mathbf{Y})>, A \in \mathcal{A}(n,m)\}.
\end{equation}
Following~\cite{ko2022video}, we define the pairwise distance as the negative cosine similarity:
\begin{equation}\label{eq:dist_matrix}
    C(\mathbf{X},\mathbf{Y}) = 1 - S(\mathbf{X},\mathbf{Y}),
\end{equation}
where $S(\mathbf{X},\mathbf{Y})$ denotes the cosine similarity between the melody and lyrics embeddings. 
This choice is motivated by our use of L2-normalized embeddings and a dot-product-based dual-encoder architecture (see Fig.~\ref{fig:MLM}).
Alternative measures of distance, e.g., Euclidean distance~\cite{tsiamas2025sequential}, are also possible.

\subsection{Length-informed Alignment Regularization}\label{sec:length_regularzation}

Although not strictly exhibiting a one-to-one correspondence, the melody and lyrics sequences of existing songs tend to have similar lengths in terms of number of notes and sylphones, respectively.
We provide statistics on sequence length differences between melody and lyrics in Section~\ref{sec:dataset}.

To make our alignment cost responsive to length differences, we introduce a regularization term to the STDW loss $D^{\gamma}(\mathbf{X}_i,\mathbf{Y}_j)$. For a given melody query $\mathbf{X}_i$ with $n_i$ notes and  a set of $B$ lyrics candidates $\mathbf{Y}_j$, each with $m_j$ sylphones, the regularized alignment cost is defined as:
\begin{equation}\label{eq:lendiff_penalise}
\tilde{D}^{\gamma}(\mathbf{X}_i,\mathbf{Y}_j) = (1 - \alpha) D^{\gamma}(\mathbf{X}_i,\mathbf{Y}_j) +  \alpha \frac{|n_i - m_j|}{\max_j^B |n_i - m_j| + \epsilon} \Delta,
\end{equation}
where $\Delta = \Big (\max_j^B D^{\gamma}(\mathbf{X}_i,\mathbf{Y}_j) - \min_j^B D^{\gamma}(\mathbf{X}_i,\mathbf{Y}_j) \Big)$,  $j=1,...B$, $\alpha$ is the regularization weight, and $\epsilon > 0$ is a small additive that avoids division by zero.
Multiplying by $\Delta$ guarantees equivalent scaling between the two right-hand side terms.
Thereafter, we refer to $\tilde{D}^{\gamma}(\mathbf{X}_i,\mathbf{Y}_j)$ as the alignment cost.

\subsection{Contrastive Alignment Loss}

We propose to apply the alignment loss $\tilde{D}^{\gamma}$ in a contrastive setting and define the \emph{contrastive alignment loss} (CAL) for MLM.
CAL is a variant of the InfoNCE loss~\cite{oord2018representation}, in which we replace the cosine similarity with the negative alignment loss:
\begin{align}\label{eq:L_cal}
    \mathcal{L} = & - \frac{1}{B}\sum_{i=1}^{B} \log \frac{ \exp{ \Big( - \tilde{D}^{\gamma}(\mathbf{X}_i, \mathbf{Y}_i) /\tau \Big) }}{\sum_{j=1}^{B} \exp{ \Big( - \tilde{D}^{\gamma}(\mathbf{X}_i, \mathbf{Y}_j)  /\tau \Big)}} \nonumber \\
    & - \frac{1}{B}\sum_{j=1}^{B} \log \frac{ \exp{ \Big( - \tilde{D}^{\gamma}(\mathbf{Y}_i, \mathbf{X}_i) /\tau \Big) }}{\sum_{i=1}^{B} \exp{ \Big( - \tilde{D}^{\gamma}(\mathbf{Y}_j, \mathbf{X}_i)  /\tau \Big)}},
\end{align}
where $\tau$ is the temperature parameter and $B$ is the batch size.
This formulation is similar to the sequential contrastive loss proposed in~\cite{tsiamas2025sequential} since $\tilde{D}^{\gamma}(\mathbf{X}_i, \mathbf{Y}_i)$ is a value computed from two sequences of embedding vectors: $\mathbf{X}_i = \bm{x}_1, ..., \bm{x}_{n}$ and $\mathbf{Y}_j = \bm{y}_1, ..., \bm{y}_{m}$.

Due to the long sequence nature of both melody and lyrics, the alignment cost $\tilde{D}^{\gamma}(\mathbf{X}_i, \mathbf{Y}_i)$  will exhibit small differences between the positive and negative pairs. 
To amplify these differences, we follow~\cite{tsiamas2025sequential} and apply batch-wise Z-score normalization to the SDTW cost:
\begin{equation}
    \tilde{D}^{\gamma}(\mathbf{X}_i, \mathbf{Y}_j) = \frac{\tilde{D}^{\gamma}(\mathbf{X}_i, \mathbf{Y}_j) - {\rm mean}_{j=1}^{B} {\tilde{D}^{\gamma}(\mathbf{X}_i, \mathbf{Y}_j)}}{{\rm std}_{j=1}^{B} {\tilde{D}^{\gamma}(\mathbf{X}_i, \mathbf{Y}_j)}},
\end{equation}
where ${\rm mean}_{j=1}^{B} {\tilde{D}^{\gamma}(\mathbf{X}_i, \mathbf{Y}_j)}$ and ${\rm std}_{j=1}^{B} {\tilde{D}^{\gamma}(\mathbf{X}_i, \mathbf{Y}_j)}$ are the mean and standard deviation of $\tilde{D}^{\gamma}(\mathbf{X}_i, \mathbf{Y}_j)$ along $j$. 
This normalization sharpens the contrast between positive and negative alignment scores. The normalized cost along $i$ can be obtained similarly.
We describe the negative sampling strategy in Section~\ref{sec:training}.

\section{Experiments}\label{sec:introduction}

\subsection{Datasets}\label{sec:dataset}
We train the proposed architecture in Fig.~\ref{fig:MLM} on the DALI V2 dataset~\cite{meseguer2020creating} which provides time-aligned vocal melody notes and lyrics of 7,756 songs at four levels of granularity: notes, words, lines, and paragraphs.
For this work, we exclude duplicate songs and restrict our analysis to English-language lyrics, comprising a subset of 5,150 songs.
Note that the annotations of the DALI dataset were created with automatic methods with a relatively low accuracy.
Importantly, our proposed method does not rely on alignment annotations, as model training is fully self-supervised. This enables the system to scale efficiently with the availability of additional songs.

Since there is no existing evaluation dataset with syllable-level annotations, we randomly select 50 songs from the DALI dataset and manually correct their annotations.
We call this subset \emph{DALI50} and use it as our evaluation dataset.
The remaining songs are regarded as development data and are randomly split into training and validation sets with a ratio of 8:2.

As the first endeavor in the MLM task, we perform melody and lyrics matching at segment level. 
This choice is driven by the practical insight that assessing the alignment between melody and lyrics does not require the full length of each modality.
Segments comprising a sufficient number of melody notes and lyrics sylphones are adequate to support the above judgment for meaningful alignment.
However, we do not fetch segments based on the amount of notes or sylphones which may not always have a one-to-one correspondence. 
Instead, we extract melody and lyrics segments with a fixed number of lines.
This approach guarantees that the segments are alignable from beginning to end.
It is important to highlight that we only use line information from melodies and lyrics (provided in the DALI dataset) for segmentation purposes. 
No such information is used during the training process.

\begin{figure}[ht]
    \centering
    \includegraphics[width=0.95\linewidth]{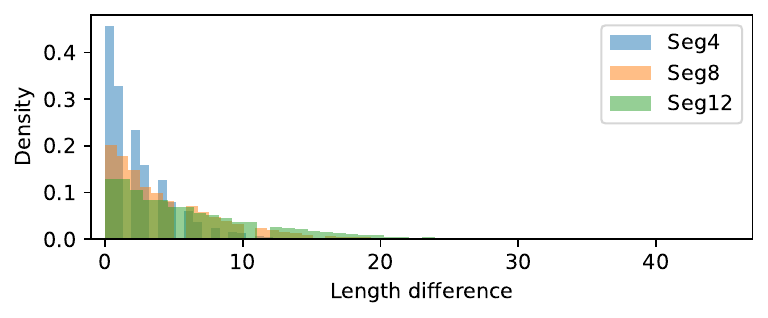}\\
    \caption{Length difference between melody-lyrics segment pairs in the development set.}
    \label{fig:length_difference}
\end{figure}

Table~\ref{tab:data_size} lists the size of each dataset subset depending on the number of lines used for segmenting. 
In this study, we consider segments composed of 4, 8, and 12 lines, denoted as \emph{Seg4}, \emph{Seg8}, and \emph{Seg12}, respectively. 
We also report the original dataset size at full-song level (in \textit{italic}), for reference.
Clearly, the number of segments decreases as the number of lines per segment increases.
To examine the length discrepancy between melody and lyrics segments, we plot histograms of the length difference of melody-lyrics pairs in the development set, as shown in Fig.~\ref{fig:length_difference}.
The histograms evidence that melody notes and lyrics sylphones do not exhibit a strict one-to-one correspondence. 
Nonetheless, the distribution displays a long-tail shape, indicating that most melody and lyrics segments have similar lengths. 
This observation motivates our introduction of the length-informed regularization term to the alignment cost in Section~\ref{sec:length_regularzation}.

\begin{table}[ht]
\centering
\caption{Dataset size with different segmentation strategies compared to the original dataset size at the full song level (in \textit{italic}).}
\label{tab:data_size}
\begin{adjustbox}{max width=\linewidth}
\rowcolors{2}{gray!10}{white}
\begin{tabular}{@{}p{1.7cm}p{1.7cm}p{1.7cm}p{1.7cm}@{}}
\toprule
\textbf{Segment} & \textbf{Training} & \textbf{Validation} & \textbf{Test} \\
\midrule
Seg4        & 37,396 & 4,274  & 572  \\
Seg8        & 18,859 & 2,162  & 308  \\
Seg12       & 11,975 & 1,352  & 203  \\
\midrule
\textit{Song} & \textit{5,150} & \textit{573} & \textit{50} \\
\bottomrule
\end{tabular}
\end{adjustbox}
\end{table}

In addition to segment length differences, we observe that both the note and sylphone vocabularies follow strong long-tail distributions. 
To facilitate model learning, we filter out low-frequency items by identifying notes and sylphones that occur fewer than 10 times in the development set. 
Subsequently, we exclude segments containing any of these rare notes and sylphones.
Due to the inherent noisiness of the DALI dataset, we further filter out segments with anomalous line lengths. 
Specifically, we retain only those segments whose melody and lyrics line lengths fall within the $5$th to $95$th percentile range of their respective distributions in the development set. This corresponds to intervals of [3, 11] for melody lines and [2, 10] for lyrics lines.

\subsection{Training and Inference}\label{sec:training}

For each encoder in Fig.~\ref{fig:MLM}, we use a 2-layer Transformer with 4 attention heads, model dimension of 256, and feed-forward dimension of 1024. 
We apply a fixed value of $\gamma=1.0$ to compute the SDTW cost across all experiments.
Our model contains around 3.7M trainable parameters in total, which is significantly smaller compared to existing lyrics generation systems \cite{ma2021ai, liangxai}. 
This compactness is attributed to the nature of the MLM task, which focuses on modeling the relationships between melody and lyrics, rather than generating either modality.
Consequently, this not only reduces overall complexity of the model but also makes it feasible to train using smaller datasets.

We set a batch size of $B=32$ melody-lyrics segment pairs for training.
We employ the Adam optimizer~\cite{loshchilov2018decoupled} coupled with a cosine learning rate scheduler and a linear warmup over the first 2 epochs.
The base learning rate is set to 
$1\mathrm{e}{-5}$ and the model is trained for 20 epochs.
To stabilize training, we apply gradient clipping with a maximum norm of 1.0.
The model checkpoint with the lowest validation loss is saved for evaluation. 
The training time varies with the size of the dataset, with larger datasets requiring longer durations. 
For example, training for the case of Seg12 requires approximately 60 hours on an NVIDIA V100 GPU. 
Note that the main computational overhead occurs during the validation phase, where each melody query needs to be aligned with all candidates in the validation set, similar to other retrieval systems.
In contrast, the training phase is relatively efficient.

During training, we create negative samples for the contrastive alignment loss by randomly shuffling both melody and lyrics.
Random shuffling breaks the note-sylphone correspondence, the structure of music, and the rhyme in lyrics.
For each melody-lyrics pair $(\mathbf{X}_i, \mathbf{Y}_i)$ in a mini-batch of size $B$, we create $B-1$ negative lyrics for each melody $\mathbf{X}_i$ by shuffling the sylphones; and similarly generates negative melodies by shuffling the notes.
Importantly, our SDTW implementation supports parallel processing of  variable-length sequences for the first time, and is built upon the framework introduced in~\cite{maghoumi2021deepnag}.

At inference time, we extract melody and lyrics embeddings using the trained dual encoder.
For alignment, we apply classical DTW to these embeddings to obtain a unique alignment path. 
To reduce the computational overhead of the retrieval process, and motivated by the comparable sequence lengths of melody and lyrics (see Section~\ref{sec:dataset}), we pre-select 50\% of candidates based on the lowest absolute length difference with the query.
Alignment costs are then computed only for this subset, and candidates are ranked accordingly.

To assess the model’s ability to retrieve meaningful lyrics from general text, we augment each lyrics candidate in the validation and test sets with a plain-text variant.
This plain version preserves the original segment length but samples sylphones randomly from the corresponding dataset.
Thus, lyrics candidates in the validation or test dataset consist of a mix of the original lyrics and their plain version.
For example, in Seg12 (see Table~\ref{tab:data_size}), each melody query has $1352 \times 2= 2704$ lyrics candidates during validation.

\subsection{Baseline and Length-informed Method}\label{sec:metrics}

As MLM is a novel task, there are no existing baselines. 
In this paper, we propose two simple baselines: a random baseline and a length-informed system.
The \textbf{random baseline} randomly selects a lyrics segment from the evaluation set and aligns it ``diagonally" with the melody.
This ``diagonal" path approximates the exact diagonal when the two sequences differ in length, which we estimate using the Bresenham's line algorithm~\cite{bresenham1965algorithm}.
The
\textbf{length-informed (LI) method} leverages segment length information as a ranking criterion. 
It ranks lyrics candidates based on the absolute difference in length between the candidates and the melody query, prioritizing those whose lengths are closest to that of the melody query.
After this ranking, the Bresenham's line algorithm is applied again to align the melody with each lyrics candidates. 
We evaluate the quality of these matching strategies, using the metrics described in Section~\ref{sec:evaluation}.

\section{Evaluation}\label{sec:evaluation}

\subsection{Objective Metrics}\label{sec:eval_obj}

Existing lyrics generation metrics~\cite{watanabe2018melody, ma2021ai} mainly assess the intelligibility of generated text, which makes them unsuitable for evaluating MLM. 
To address this, we propose a novel set of objective metrics to measure the matching quality in MLM. These include the Hit@K metric, which evaluates the retrieval aspect, and seven additional metrics that measure the alignment quality.

\vspace{.5ex}
\noindent \textbf{Hit@K}: This is a standard retrieval metric defined as the ratio of melody queries that are successfully matched to the original lyrics within the top-K\% of retrieved lyrics, relative to the total number of queries.
It is based on the assumption that the reference (original lyrics) generally matches well with the melody query. 
We employ a top-K\% selection rather than using a fixed top-K, as the number of lyrics candidates varies depending on the number of lines used during the segmentation of the lyrics.

\vspace{.5ex}
\noindent \textbf{Stress matching rate (SMR)}: Prior studies have demonstrated a statistical correlation between musical and lyrical stresses~\cite{nichols2009relationships}, e.g., there is a correspondence between long notes and high stress levels, long vowels, and non-stopwords.
SMR is akin to the prominent
word-note matching rate metric in~\cite{zhao2025reffly}.
In this work, we define long notes as those whose durations exceed the third quartile of the note duration distribution in the song. 
Among the sylphones matched to these long notes, we compute the percentage of long vowels, stress level of at least 1, and non-stopwords.
Long vowels include AA, AO, AW, AY, EY, IY, OW, OY, UW, following~\cite{nichols2009relationships}.
We regard words that are not in the stopword list of the \emph{nltk} library~\cite{bird2009natural} as non-stopwords.

\vspace{.5ex}
\noindent \textbf{Rhyme density/distance/strength (R3)}: 
For lyrics, rhyme often appears at the last syllable of each line \cite{pattison1991songwriting}. 
As presented in Section~\ref{sec:sylphone},  the sylphone representation is particularly effective for capturing rhyme information through its rhyming elements, specifically the vowel and end consonants.
We use line information during the evaluation stage to compute rhyme metrics.

For a song with $L$ lyric lines, let $r_i$ denote the rhyming elements of the last sylphone of the $i$-th line, and $r_i^v$ as its vowel component. 
Similarly, $\hat{r}_i$ and $\hat{r}_i^v$ represent those of the matched lyrics.  
We identify rhyming positions $i_k$ as those where the same vowel appears at least twice across line endings, where $k=1,...K$.
Rhyme density quantifies the frequency of rhyming within the lyrics and is defined as the ratio of repeated vowels to the total number of lines, expressed as $R_{\rm den} = K/L$.

Ideally, rhyme strength measures the degree of phonological similarity between rhyming elements, taking into account both vowel and consonant features. 
A higher degree of phonological similarity indicates a stronger rhyme. 
In this work, we consider a simplified version of this metric. We assess whether the rhyming sylphones share the same vowel, and, in cases where the vowels match, whether the associated consonants belong to the same category.
For the rhyming sylphones $r_{i_k}$, we compute the number of unique vowels $u = |\{r_{i_k}^v\}|$.
For sylphones sharing the same vowel, $r_{i_s}$, where $s=1,...,S$, we obtain the number of unique end consonants $w = |\{r_{i_s}^v\}|$. 
The overall rhyme strength is defined as $R_{\rm str} = \frac{1}{2}(\frac{u}{K} + \frac{w}{S})$.

Rhyme density and strength are metrics  that serve to quantify the rhyme characteristics within lyrics.
To evaluate how well the rhyme structure of the matched lyrics aligns with that of the original lyrics, we introduce a rhyme distance metric.
We encode the rhyme positions into a multi-hot vector $p \in \{0,1\}^L$ and define rhyme distance between the matched lyrics $\hat{p}$ and the original lyrics $p$ as $R_{\rm dis} = \frac{\|\hat{p} - p\|}{\sum_i^L\hat{p}\cup p}$.

\vspace{.5ex}
\noindent \textbf{Frequency of extreme matches (FEM)}: 
Although a strict one-to-one correspondence between note and sylphone is not always observed, occurrences of substantial local warping, i.e. mapping multiple sylphones to a single note or vice versa, is relatively rare.
Such cases would result in lyrics that are difficult or impossible to sing.
To measure the frequency of these extreme matches, we compute the mean of the maximum number of notes matched to a single sylphone and vice versa.

\subsection{Results}\label{sec:introduction}

\textbf{Retrieval evaluation}: 
Table~\ref{tab:result_table} displays the Hit@K metric values obtained from the random baseline, length-informed~(LI), and our proposed MLM-CAL method on the evaluation set with different segmentation strategies. 
Comparing the results across methods, the length-informed method significantly outperforms the random baseline for all $K$s and all segmentation cases.
Note that our test set includes the original DALI50 and a plain-text version of it obtained by random sampling sylphones.
This is why the performance of the random baseline is $K/2$.
MLM-CAL achieves the best performance, with a large performance boost in all cases compared to the length-informed method. 

\begin{figure*}[t]
    \centering
    \includegraphics[width=\linewidth]{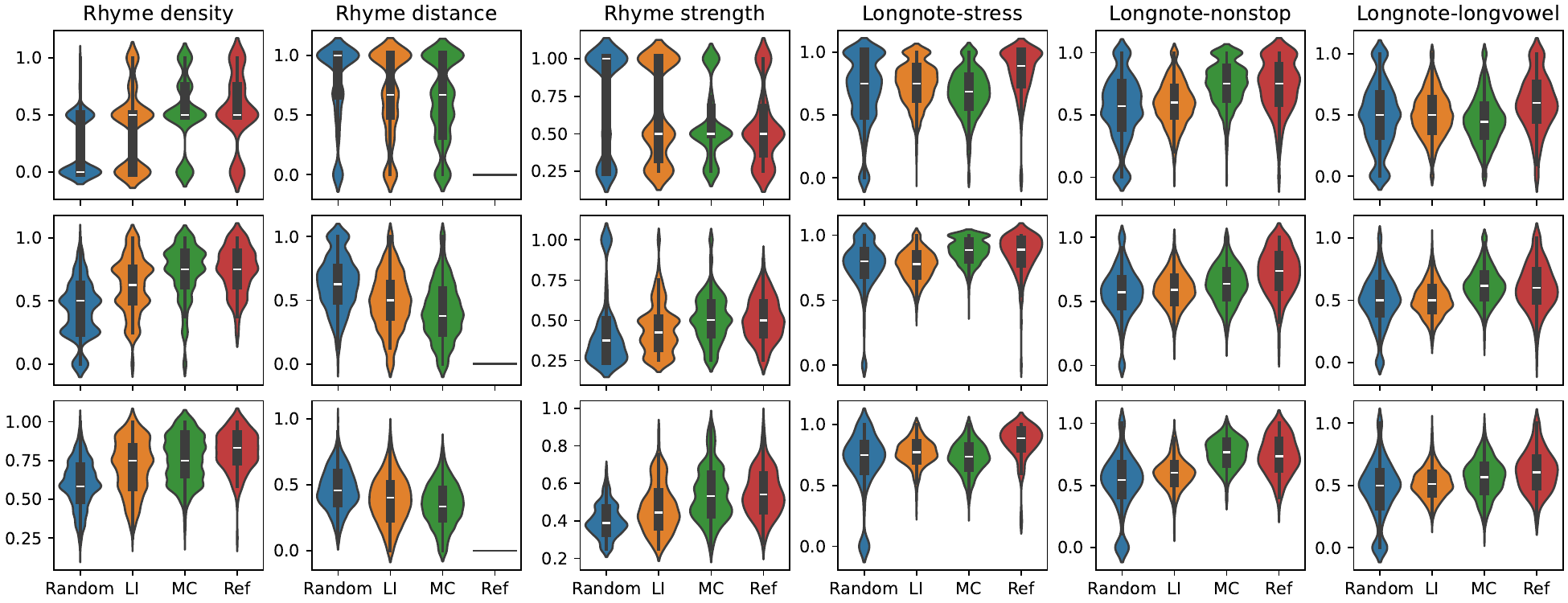}
    \caption{Violin plots of the average matching quality metrics over the top 5 matched lyrics for Seg4 (top), Seg8 (middle), and Seg12 (bottom). 
    \textbf{Random, LI, MC,} and \textbf{Ref} refer to the random baseline, length-informed method, MLM-CAL, and reference, respectively. Each column corresponds to a specific metric (from left to right): rhyme density, rhyme strength, rhyme diversity, longnote-stress, longnote-nonstop, and longnote-longvowel correspondence.}
    \label{fig:align_quality}
\end{figure*}

\begin{table}[ht]
\centering
\caption{Hit@K\% on the evaluation set from the random baseline, length-informed, and our proposed MLM-CAL method. NA means ``not applicable". The best results are shown in \textbf{bold} and the second-best are \underline{underlined}.}
\label{tab:result_table}
\begin{adjustbox}{max width=\linewidth}
\rowcolors{2}{gray!10}{white}
\begin{tabular}{@{}p{2cm}p{1.2cm}p{0.6cm}p{1.1cm}p{1.1cm}p{1.1cm}@{}}
    \toprule
    \textbf{Method} & \textbf{Segment} & \textbf{$\alpha$} & \textbf{Hit@1\%} & \textbf{Hit@3\%} & \textbf{Hit@5\%}\\ \midrule
    Random baseline      & Seg4/8/12  & NA & 0.5 & 1.5  & 2.5\\ \midrule
    Length-informed           & Seg4 & NA & 1.92 & 10.84 & 25.87\\
    MLM-CAL     & Seg4 & 0.25 & 9.62 & 22.73 & 33.39\\
    MLM-CAL     & Seg4 & 0.5 & \underline{16.78} & \textbf{39.16} & \textbf{51.57}\\ 
    MLM-CAL     & Seg4 & 0.75 & \textbf{18.71} & \underline{38.46} & \underline{45.45}\\ 
 \midrule
    Length-informed & Seg8 & NA & 2.60 & 15.26 & 27.27 \\
    MLM-CAL     & Seg8 & 0.25 & 4.22 & 14.61 & 19.48\\
    MLM-CAL     & Seg8 & 0.5 & \textbf{14.29} & \underline{22.40} & \underline{28.90}\\ 
    MLM-CAL     & Seg8 & 0.75 & 11.69 & \textbf{25.97} & \textbf{33.77}\\
 \midrule
    Length-informed & Seg12& NA  & 3.94 & 15.76 & 24.14 \\
    MLM-CAL     & Seg12 & 0.25 & 8.87 & 20.69 & 25.62\\
    MLM-CAL    & Seg12 & 0.5 & \underline{10.84} & \underline{22.17} & \underline{32.02}\\
    MLM-CAL     & Seg12 & 0.75 & \textbf{11.33} & \textbf{25.12} & \textbf{32.51}\\
    \bottomrule
\end{tabular}
\end{adjustbox}
\end{table}

\begin{figure}[t]
    \centering
    \includegraphics[width=\linewidth]{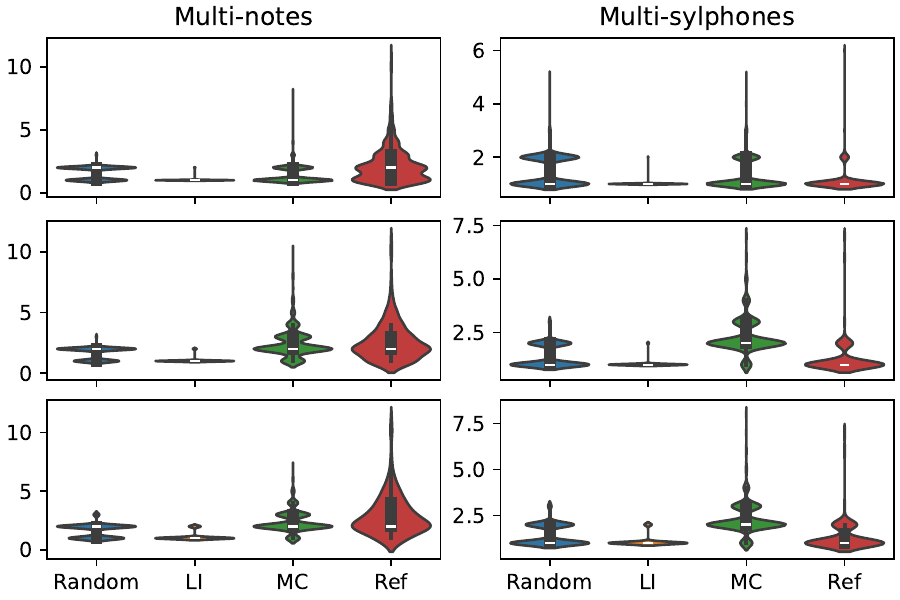}
    \caption{Violin plots of the average FEM metrics on multi-notes (left) and multi-sylphones (right) over the top 5 matched lyrics for Seg4 (top), Seg8 (middle), and Seg12 (bottom). }
    \label{fig:extreme_macthes}
\end{figure}

When extending the comparison to different segment lengths, i.e., Seg4, Seg8, and Seg12, we observe performance drops for MLM-CAL as segment length increases. 
Ideally, longer segments allow the proposed method to exploit richer information, including rhyme patterns and structural alignment between the melody and lyrics, beyond local correspondence. 
In contrast, this advantage is less pronounced for MLM-CAL in the Seg12 setting, likely due to the limited amount of training data available for longer segments.
Indeed, the quantity of training data for Seg12 is substantially lower compared to that for Seg4 and Seg8 (see Table~\ref{tab:data_size}).

To investigate the impact of the regularization term in Equation~\ref{eq:lendiff_penalise}, we perform an ablation study on the weight parameter, with $\alpha \in \{0.25, 0.5, 0.75\}$, as shown in Table~\ref{tab:result_table}.
The results indicate that longer segments benefit more from stronger regularization. 
For example, in the case of Seg12, increasing $\alpha$ from $0.5$ to $0.75$ leads to consistent improvements in Hit@K metrics. In contrast, for Seg4, only Hit@1\% shows improvement, while Hit@5\% experiences a substantial drop.
A notable performance degradation is observed for $\alpha = 0.25$ across all cases.

\begin{figure*}[ht]
    \centering
    \includegraphics[width=\linewidth]{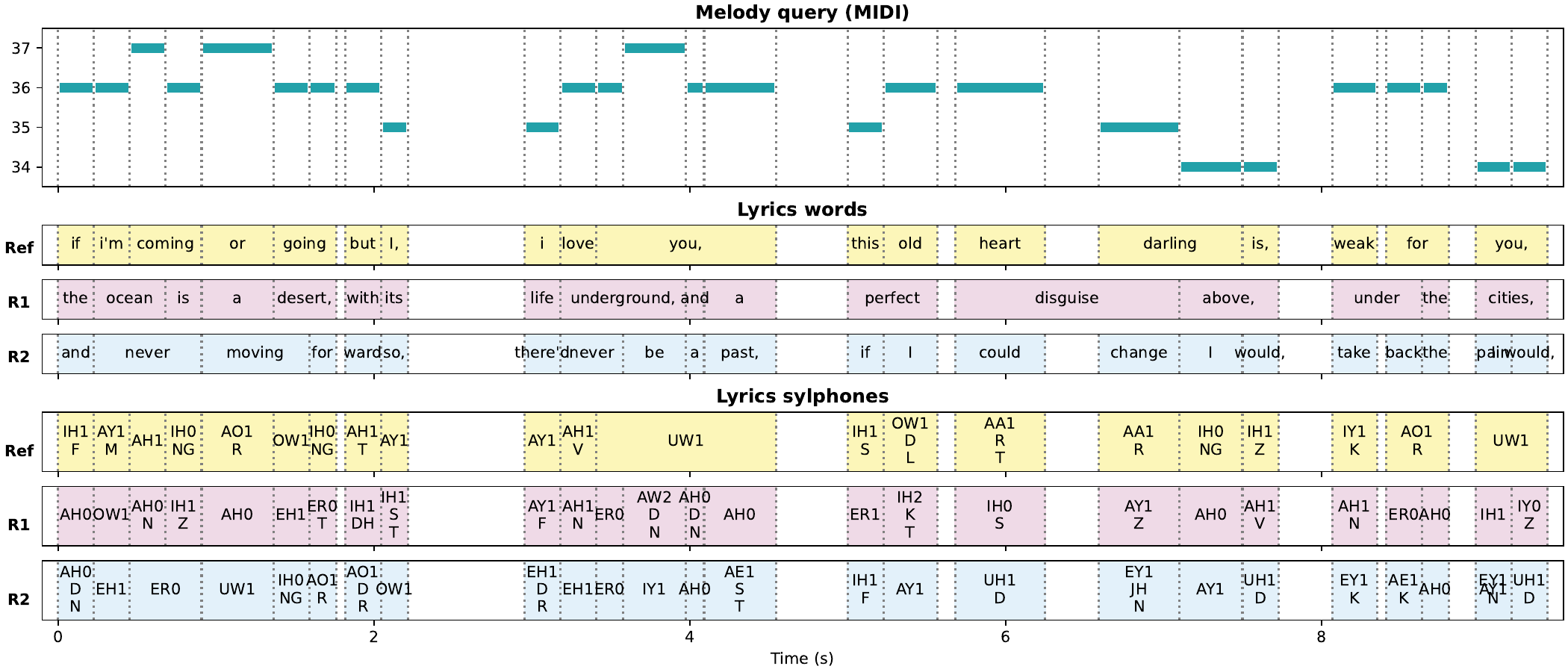}
    \caption{Alignment visualization of top retrieved lyrics for a melody query from the song with ID \emph{0c309f7a0d31428da09e71d645426963} in DALI50. Top: melody query in MIDI notes. Middle: lyrics words of the reference (\textbf{Ref}), rank 1 (\textbf{R1}), and rank 2 (\textbf{R2}) lyrics. Bottom: lyrics sylphones of the reference, rank 1, and rank 2 lyrics. We provide more examples with melody playback on our companion webpage.}
    \label{fig:example_match}
\end{figure*}

\vspace{0.5ex}
\textbf{Alignment quality}: Fig.~\ref{fig:align_quality} shows the violin plots of the alignment quality metrics on the evaluation set, obtained from models trained with $\alpha=0.5$.
Each row corresponds to a specific segmentation setting, with Seg4, Seg8, and Seg12 arranged from top to bottom.
Each column represents a different metric, including the R3 metrics (rhyme density, distance, and strength) and the SMR metrics (longnote-stress, longnote-nonstop, and longnote-longvowel correspondence).

The rhyming metrics primarily reflect long-range structural correspondence. 
A consistent performance hierarchy is observed across all methods: reference $>$ MLM-CAL $>$ length-informed $>$ random baseline. 
This ranking becomes increasingly pronounced with longer segments, as MLM-CAL better exploits higher-level features such as rhyme.

The disparity between methods in the rhyme-metrics is especially prominent for Seg12 and Seg8, whereas the differences for Seg4 are hardly observed. 
This is because rhyme patterns are less likely to emerge in shorter sequences (e.g., 4 lines). 
It is possible that there is no rhyme in a 4-line segment.
These observations evidence that
the longer the sequence, the greater the potential of MLM-CAL, if more data is available.

The stress matching rate metrics emphasize local correspondence at the note-sylphone level.
Among these metrics, the most salient feature used by the model appears to be the alignment between long notes and non-stopwords. 
In contrast, the differences across methods in the longnote–stress and longnote–longvowel correspondences are less obvious. 
A potential explanation lies in the broad and imbalanced distributions of both stress and longvowels features.
As described in Section~\ref{sec:eval_obj}, we compute the proportion of long notes that align with stress levels of at least 1. However, given that only three stress levels are defined and level 2 is infrequently observed, this limits the discriminative power of stress-related metrics.
Similarly, for the longnote-longvowel correspondence, we compute the ratio of long notes aligned with vowels in the long vowel list, which accounts for more than half of all vowels. 

The frequency of extreme matches is analyzed primarily as a sanity check, ensuring that no anomalous alignments occur, such as a single note being matched to a large number of sylphones or vice versa. 
To visualize this, we plot the distribution of extreme matches for all aforementioned methods using violin plots, as shown in Fig.~\ref{fig:extreme_macthes}.
The observed distribution of the reference confirms that note-sylphone correspondences do not strictly follow a one-to-one mapping. 
One-to-few correspondences are relatively common, particularly in the form of ``multi-note” cases, where a single sylphone is aligned to multiple notes.

When comparing across methods, the proposed approach exhibits a distribution that most closely resembles the reference, indicating a more natural alignment behavior.
In contrast, the length-informed method tends to center sharply around a one-to-one correspondence in both directions. 
This is expected, as this method aligns based on the closest or exactly equal sequence lengths, resulting in a near-diagonal alignment. 
However, this strategy fails to capture realistic alignment dynamics in read-world songs.
The random baseline exhibits a broader distribution due to the random selection from candidates.

\vspace{0.5ex}
\textbf{Alignment visualization}
Fig.~\ref{fig:example_match} displays an example of the rank 1 and rank 2 lyrics retrieved by our proposed MLM-CAL method, compared to the reference lyrics aligned with the melody query.
As can be seen, MLM-CAL provides different possibilities of lyrics, which may differ semantically from the reference but still align with the same melody.
Take the second line, for instance: our rank 2 candidate matches \textit{``there'd never be a past"} to the corresponding melody, replacing the reference lyrics \textit{``I love you"}.
The former is 6-7 note-sylphone alignment, while the latter is 6-3, especially as \textit{``you"} is aligned to 4 notes (including 2 long ones).

Checking the local correspondence within lines, we confirm that in most cases, MLM-CAL indeed matches short notes with stopwords, such as \textit{``a"} and \textit{``for"}, whereas long notes are aligned to non-stopwords. 
Expanding the analysis window more broadly to the whole segment, we observe that our proposed system currently struggles with aligning melody and lyrics line-by-line; e.g., the ends of the first two lines (indicated by ``,") of the rank 1 lyrics are not aligned with the melody.
In fact, we do not have any explicit information on melody or lyrics lines during training.
Besides lyrics words, we also provide alignment visualization of lyrics sylphones.
The note-sylphone alignment is especially useful for singing in polysyllabic languages such as English.
We provide more examples with melody playback on our companion webpage\footnote{https://github.com/changhongw/mlm}.

\subsection{Discussion}
\subsubsection{Large potential for retrieving lyrics from general text} 
MLM is a challenging task.
The visualization in Fig.~\ref{fig:example_match} demonstrates the complexity of balancing multiple alignment criteria when retrieving suitable lyrics from existing text. 
The system must not only ensure line-level matches but also capture fine-grained features such as local stress patterns.
When the search database is limited, such as in our evaluation setting with only 50 candidate songs, it becomes difficult to retrieve high-quality lyrics. 
This constraint defines the upper bound of achievable performance within the current dataset. 
Nevertheless, the proposed method has strong potential for generalization. 
In particular, it can be extended to retrieve lyrics from much broader sources of text, such as large-scale corpora, which are virtually unlimited in size. 

\subsubsection{Scalability Through Incorporation of Additional Data}
The proposed method is fully self-supervised and does not rely on any manual alignment annotations, making it inherently scalable with the availability of additional data. 
As more songs become accessible, the system can be extended seamlessly without requiring changes to the training paradigm.
Furthermore, while this study focuses on symbolic melodies as a starting point, the framework is not limited to this representation. With appropriate modifications to the melody encoder, the method can potentially be adapted for use with raw audio inputs, enabling broader applicability across different music formats and datasets.

\subsubsection{Retrieving melody for lyrics or poems}
While the focus of this paper is on retrieving lyrics for a given melody query, MLM-CAL is inherently bidirectional and can be extended to retrieve melodies for a given lyrical text, such as existing lyrics or poems. 
This flexibility arises from the symmetric nature of the training objective, as described in Equation~\ref{eq:L_cal}, and the bidirectional architecture of the model illustrated in Fig.~\ref{fig:MLM}. 
As a result, the framework holds potential for broader applications, including text-to-melody retrieval tasks so that one can sing along the same lyrics with a variety of melodies.


\subsubsection{Limitations}
In addition to the limited size of the training dataset, we acknowledge three other limitations in the current pipeline.
First, we only considered within-domain evaluation where the evaluation songs are also from the DALI dataset. 
Second, the evaluation is based solely on objective metrics. 
Although these metrics provide quantifiable insights into system performance, they do not capture perceptual or qualitative aspects of alignment quality. A more comprehensive evaluation, including human judgments and subjective assessments, is planned for future work.
Finally, with respect to stress correspondence, our current approach models this aspect only implicitly via the relationship between long notes in the melody and stress patterns in the lyrics. 
Incorporating more explicit musical features, such as beats and downbeats, could potentially enhance stress alignment performance. However, the current dataset does not include such annotations.

\section{Conclusion}
We propose melody-lyrics matching, a new way of pairing melody and lyrics by retrieving potential lyrics for a melody query from existing text databases. 
It provides different possibilities of lyrics in addition to the original ones.
Approaching this task as a self-supervised representation problem, we propose to learn the embeddings of melody and lyrics via a contrastive alignment loss specifically designed to capture the sequential nature of music. 
A promising avenue is to explore the scalability of this approach when applied to large, open-domain text corpora, such as general web text, for lyric retrieval, where data availability is virtually unlimited.

\bibliographystyle{IEEEtran}
\bibliography{bibliography}

\begin{thebibliography}{10}
\providecommand{\url}[1]{#1}
\csname url@samestyle\endcsname
\providecommand{\newblock}{\relax}
\providecommand{\bibinfo}[2]{#2}
\providecommand{\BIBentrySTDinterwordspacing}{\spaceskip=0pt\relax}
\providecommand{\BIBentryALTinterwordstretchfactor}{4}
\providecommand{\BIBentryALTinterwordspacing}{\spaceskip=\fontdimen2\font plus
\BIBentryALTinterwordstretchfactor\fontdimen3\font minus \fontdimen4\font\relax}
\providecommand{\BIBforeignlanguage}[2]{{%
\expandafter\ifx\csname l@#1\endcsname\relax
\typeout{** WARNING: IEEEtran.bst: No hyphenation pattern has been}%
\typeout{** loaded for the language `#1'. Using the pattern for}%
\typeout{** the default language instead.}%
\else
\language=\csname l@#1\endcsname
\fi
#2}}
\providecommand{\BIBdecl}{\relax}
\BIBdecl

\bibitem{watanabe2018melody}
K.~Watanabe, Y.~Matsubayashi, S.~Fukayama, M.~Goto, K.~Inui, and T.~Nakano, ``A melody-conditioned lyrics language model,'' in \emph{Proceedings of the 2018 Conference of the North American Chapter of the Association for Computational Linguistics: Human Language Technologies, Volume 1 (Long Papers)}, 2018, pp. 163--172.

\bibitem{ma2021ai}
X.~Ma, Y.~Wang, M.-Y. Kan, and W.~S. Lee, ``Ai-lyricist: Generating music and vocabulary constrained lyrics,'' in \emph{Proceedings of the 29th ACM International Conference on Multimedia}, 2021, pp. 1002--1011.

\bibitem{pan2022vocal}
Y.~Pan, C.~Landreth, E.~Fiume, and K.~Singh, ``Vocal: Vowel and consonant layering for expressive animator-centric singing animation,'' in \emph{SIGGRAPH Asia 2022 Conference}, 2022, pp. 1--9.

\bibitem{panteli2017towards}
M.~Panteli, R.~Bittner, J.~P. Bello, and S.~Dixon, ``Towards the characterization of singing styles in world music,'' in \emph{IEEE International Conference on Acoustics, Speech and Signal Processing (ICASSP)}.\hskip 1em plus 0.5em minus 0.4em\relax IEEE, 2017, pp. 636--640.

\bibitem{kruspe2016retrieval}
A.~M. Kruspe and I.~Fraunhofer, ``Retrieval of textual song lyrics from sung inputs.'' in \emph{INTERSPEECH}, 2016, pp. 2140--2144.

\bibitem{durand2023contrastive}
S.~Durand, D.~Stoller, and S.~Ewert, ``Contrastive learning-based audio to lyrics alignment for multiple languages,'' in \emph{IEEE International Conference on Acoustics, Speech and Signal Processing (ICASSP)}, 2023, pp. 1--5.

\bibitem{schulze2021phoneme}
K.~Schulze-Forster, C.~S. Doire, G.~Richard, and R.~Badeau, ``Phoneme level lyrics alignment and text-informed singing voice separation,'' \emph{IEEE/ACM Transactions on Audio, Speech, and Language Processing (TASLP)}, vol.~29, pp. 2382--2395, 2021.

\bibitem{vaglio2020multilingual}
A.~Vaglio, R.~Hennequin, M.~Moussallam, G.~Richard, and F.~d'Alch{\'e} Buc, ``Multilingual lyrics-to-audio alignment,'' in \emph{International Society for Music Information Retrieval Conference (ISMIR)}, 2020.

\bibitem{pretet2022video}
L.~Pr{\'e}tet, G.~Richard, C.~Souchier, and G.~Peeters, ``Video-to-music recommendation using temporal alignment of segments,'' \emph{IEEE Transactions on Multimedia}, vol.~25, pp. 2898--2911, 2022.

\bibitem{komatsu2025aligned}
T.~Komatsu, H.~Munakata, T.~Hasumi, and Y.~Fujita, ``Aligned contrastive learning for text-to-music retrieval,'' in \emph{IEEE International Conference on Acoustics, Speech and Signal Processing (ICASSP)}.\hskip 1em plus 0.5em minus 0.4em\relax IEEE, 2025, pp. 1--5.

\bibitem{dave2025sync}
I.~R. Dave, F.~C. Heilbron, M.~Shah, and S.~Jenni, ``Sync from the sea: retrieving alignable videos from large-scale datasets,'' in \emph{European Conference on Computer Vision (ECCV)}, 2025, pp. 371--388.

\bibitem{tsiamas2025sequential}
I.~Tsiamas, S.~Pascual, C.~Yeh, and J.~Serr{\`a}, ``Sequential contrastive audio-visual learning,'' in \emph{IEEE International Conference on Acoustics, Speech and Signal Processing (ICASSP)}.\hskip 1em plus 0.5em minus 0.4em\relax IEEE, 2025, pp. 1--5.

\bibitem{ko2022video}
D.~Ko, J.~Choi, J.~Ko, S.~Noh, K.-W. On, E.-S. Kim, and H.~J. Kim, ``Video-text representation learning via differentiable weak temporal alignment,'' in \emph{IEEE/CVF Conference on Computer Vision and Pattern Recognition (CVPR)}, 2022, pp. 5016--5025.

\bibitem{wu2025flam}
Y.~Wu, C.~Tsirigotis, K.~Chen, C.-Z.~A. Huang, A.~Courville, O.~Nieto, P.~Seetharaman, and J.~Salamon, ``Flam: Frame-wise language-audio modeling,'' \emph{arXiv preprint arXiv:2505.05335}, 2025.

\bibitem{cuturi2017soft}
M.~Cuturi and M.~Blondel, ``Soft-{DTW}: a differentiable loss function for time-series,'' in \emph{International Conference on Machine Learning (ICML)}, 2017, pp. 894--903.

\bibitem{muller2015fundamentals}
M.~M{\"u}ller, \emph{Fundamentals of music processing: Audio, analysis, algorithms, applications}.\hskip 1em plus 0.5em minus 0.4em\relax Springer, 2015, vol.~5.

\bibitem{elizalde2023clap}
B.~Elizalde, S.~Deshmukh, M.~Al~Ismail, and H.~Wang, ``Clap learning audio concepts from natural language supervision,'' in \emph{IEEE International Conference on Acoustics, Speech and Signal Processing (ICASSP)}.\hskip 1em plus 0.5em minus 0.4em\relax IEEE, 2023, pp. 1--5.

\bibitem{nichols2009relationships}
E.~Nichols, D.~Morris, S.~Basu, and C.~Raphael, ``Relationships between lyrics and melody in popular music,'' in \emph{International Society for Music Information Retrieval Conference (ISMIR)}, 2009, pp. 471--476.

\bibitem{liu2022performance}
L.~Liu, Q.~Kong, G.~Morfi, E.~Benetos \emph{et~al.}, ``Performance midi-to-score conversion by neural beat tracking,'' in \emph{International Society for Music Information Retrieval (ISMIR)}, 2022.

\bibitem{liangxai}
Q.~Liang, X.~Ma, F.~Doshi-Velez, B.~Lim, and Y.~Wang, ``Xai-lyricist: Improving the singability of ai-generated lyrics with prosody explanations,'' in \emph{International Joint Conference on Artificial Intelligence (IJCAI) Special Track on Human-Centred AI}, 2024, pp. 7877--7885.

\bibitem{park2024real}
J.~Park, S.~Yong, T.~Kwon, and J.~Nam, ``A real-time lyrics alignment system using chroma and phonetic features for classical vocal performance,'' in \emph{IEEE International Conference on Acoustics, Speech and Signal Processing (ICASSP)}.\hskip 1em plus 0.5em minus 0.4em\relax IEEE, 2024, pp. 1371--1375.

\bibitem{low2003singable}
P.~Low, ``Singable translations of songs,'' \emph{Perspectives: Studies in Translatology}, vol.~11, no.~2, pp. 87--103, 2003.

\bibitem{antonisen2024polysinger}
S.~Antonisen and I.~L{\'o}pez-Espejo, ``Polysinger: Singing-voice to singing-voice translation from english to japanese,'' in \emph{International Society for Music Information Retrieval (ISMIR)}, 2024.

\bibitem{pattison1991songwriting}
P.~Pattison, \emph{Songwriting: Essential guide to lyric form and structure: Tools and techniques for writing better lyrics}.\hskip 1em plus 0.5em minus 0.4em\relax Hal Leonard Corporation, 1991.

\bibitem{bird2009natural}
S.~Bird, E.~Klein, and E.~Loper, \emph{Natural language processing with Python: analyzing text with the natural language toolkit}.\hskip 1em plus 0.5em minus 0.4em\relax O'Reilly Media, Inc., 2009.

\bibitem{oord2018representation}
A.~v.~d. Oord, Y.~Li, and O.~Vinyals, ``Representation learning with contrastive predictive coding,'' \emph{arXiv preprint arXiv:1807.03748}, 2018.

\bibitem{meseguer2020creating}
G.~Meseguer-Brocal, A.~Cohen-Hadria, and G.~Peeters, ``Creating {DALI}, a large dataset of synchronized audio, lyrics, and notes,'' \emph{Transactions of the International Society for Music Information Retrieval (TISMIR)}, vol.~3, no.~1, 2020.

\bibitem{loshchilov2018decoupled}
I.~Loshchilov and F.~Hutter, ``Decoupled weight decay regularization,'' in \emph{International Conference on Learning Representations (ICLR)}, 2018.

\bibitem{maghoumi2021deepnag}
M.~Maghoumi, E.~M. Taranta, and J.~LaViola, ``Deepnag: Deep non-adversarial gesture generation,'' in \emph{Proceedings of the 26th International Conference on Intelligent User Interfaces}, 2021, pp. 213--223.

\bibitem{bresenham1965algorithm}
J.~E. Bresenham, ``Algorithm for computer control of a digital plotter,'' \emph{IBM Systems Journal}, vol.~4, no.~1, pp. 25--30, 1965.

\bibitem{zhao2025reffly}
S.~Zhao, B.~Li, Y.~Tian, and N.~Peng, ``Reffly: Melody-constrained lyrics editing model,'' in \emph{Proceedings of the 2025 Conference of the Nations of the Americas Chapter of the Association for Computational Linguistics: Human Language Technologies (Volume 1: Long Papers)}, 2025, pp. 11\,295--11\,315.

\end{thebibliography}

\newpage

\end{document}